\documentclass[12pt,twoside]{article}
\usepackage[T1]{fontenc}
\usepackage[latin1]{inputenc}
\usepackage{times}
\usepackage{graphicx}
\usepackage{a4wide}
\usepackage{listings}
 \usepackage{amsmath}

\begin{document}

\section*{Future perspectives at SIS-100 with HADES-at-FAIR\footnote{Invited contribution presented at the 
    XLVII International Winter Meeting on Nuclear Physics, Bormio (Italy), Jan. 26-30, 2009}}

\thispagestyle{empty}

\begin{raggedright}

\markboth{I.Fr\"ohlich {\it et al.}}
{Future perspectives at SIS-100 with HADES-at-FAIR}

I.~Fr\"{o}hlich$^{7}$, 
G.~Agakishiev$^{8}$, A.~Balanda$^{3,e}$, G.~Bellia$^{1,a}$, D.~Belver$^{15}$,
A.V.~Belyaev$^{6}$, A.~Blanco$^{2}$, M.~B\"{o}hmer$^{11}$, J.~L.~Boyard$^{13}$, P.~Braun-Munzinger$^{4}$
% ...

\begin{center} (HADES collaboration)
\end{center}

\hspace{-0.4cm}\makebox[0.3cm][r]{$^{1}$}
Istituto Nazionale di Fisica Nucleare - Laboratori Nazionali del Sud, 95125~Catania, Italy\\
\hspace{-0.4cm}\makebox[0.3cm][r]{$^{2}$}
LIP-Laborat\'{o}rio de Instrumenta\c{c}\~{a}o e F\'{\i}sica Experimental de Part\'{\i}culas, 3004-516~Coimbra, Portugal\\
\hspace{-0.4cm}\makebox[0.3cm][r]{$^{3}$}
Smoluchowski Institute of Physics, Jagiellonian University of Cracow, 30-059~Krak\'{o}w, Poland\\
\hspace{-0.4cm}\makebox[0.3cm][r]{$^{4}$}
Gesellschaft f\"{u}r Schwerionenforschung mbH, 64291~Darmstadt, Germany\\
\hspace{-0.4cm}\makebox[0.3cm][r]{$^{5}$}
Institut f\"{u}r Strahlenphysik, Forschungszentrum Dresden-Rossendorf, 01314~Dresden, Germany\\
\hspace{-0.4cm}\makebox[0.3cm][r]{$^{6}$}
Joint Institute of Nuclear Research, 141980~Dubna, Russia\\
\hspace{-0.4cm}\makebox[0.3cm][r]{$^{7}$}
Institut f\"{u}r Kernphysik, Johann Wolfgang Goethe-Universit\"{a}t, 60438 ~Frankfurt, Germany\\
\hspace{-0.4cm}\makebox[0.3cm][r]{$^{8}$}
II.Physikalisches Institut, Justus Liebig Universit\"{a}t Giessen, 35392~Giessen, Germany\\
\hspace{-0.4cm}\makebox[0.3cm][r]{$^{9}$}
Istituto Nazionale di Fisica Nucleare, Sezione di Milano, 20133~Milano, Italy\\
\hspace{-0.4cm}\makebox[0.3cm][r]{$^{10}$}
Institute for Nuclear Research, Russian Academy of Science, 117312~Moscow, Russia\\
\hspace{-0.4cm}\makebox[0.3cm][r]{$^{11}$}
Physik Department E12, Technische Universit\"{a}t M\"{u}nchen, 85748~M\"{u}nchen, Germany\\
\hspace{-0.4cm}\makebox[0.3cm][r]{$^{12}$}
Department of Physics, University of Cyprus, 1678~Nicosia, Cyprus\\
\hspace{-0.4cm}\makebox[0.3cm][r]{$^{13}$}
Institut de Physique Nucl\'{e}aire d'Orsay, CNRS/IN2P3, 91406~Orsay Cedex, France\\
\hspace{-0.4cm}\makebox[0.3cm][r]{$^{14}$}
Nuclear Physics Institute, Academy of Sciences of Czech Republic, 25068~Rez, Czech Republic\\
\hspace{-0.4cm}\makebox[0.3cm][r]{$^{15}$}
Departamento de F\'{\i}sica de Part\'{\i}culas, University of Santiago de Compostela, 15782~Santiago de Compostela, Spain\\
\hspace{-0.4cm}\makebox[0.3cm][r]{$^{16}$}
Instituto de F\'{\i}sica Corpuscular, Universidad de Valencia-CSIC, 46971~Valencia, Spain\\ 
\hspace{-0.4cm}\makebox[0.3cm][r]{$^{a}$}
{Also at Dipartimento di Fisica e Astronomia, Universit\`{a} di Catania, 95125~Catania, Italy}\\
\hspace{-0.4cm}\makebox[0.3cm][r]{$^{b}$}
{Also at ISEC Coimbra, ~Coimbra, Portugal}\\
\hspace{-0.4cm}\makebox[0.3cm][r]{$^{c}$}
{Also at Technische Universit\"{a}t Dresden, 01062~Dresden, Germany}\\
\hspace{-0.4cm}\makebox[0.3cm][r]{$^{d}$}
{Also at Dipartimento di Fisica, Universit\`{a} di Milano, 20133~Milano, Italy}\\
\hspace{-0.4cm}\makebox[0.3cm][r]{$^{e}$}
{Also at Panstwowa Wyzsza Szkola Zawodowa, 33-300~Nowy Sacz, Poland}

\end{raggedright}
\vspace{1cm}
\begin{center}
In remembrance of our collaborators I. Iori $(\dag)$ and H. Bokemeyer $(\dag)$.
\end{center}
\vspace{0.5cm}

\begin{center}
\textbf{Abstract}
\end{center}
  Currently, the HADES spectrometer undergoes un upgrade program
to be prepared for measurements at the upcoming SIS-100 synchrotron at
FAIR.
We describe the current status of the HADES di-electron measurements at
the SIS-18 and our future plans for SIS-100.

\section{Current experimental status}

The experimental determination of hadron properties inside strongly
interacting media (normal nuclear or hot and dense matter) is one of
the very interesting challenges in hadronic physics
(see~\cite{Hayano:2008vn} for a review). As the strong coupling
$\alpha_s$ becomes very large below the QCD
scale $\Lambda_{QCD} \approx 200$~MeV, non-perturbative effects as
confinement and the breaking of chiral symmetry rule the nature of
strong interactions in the universe (beside extreme cases like neutron
stars) solely. Experimentally, the approach to get a deeper
understanding of these features is to create nuclear matter under extreme
conditions in the laboratory which can be done by employing heavy ion
reactions, only.

During the evolution of the fireball, created in the course of
heavy ion collisions,
the properties
of nuclear matter change drastically (see~\cite{BraunMunzinger:2008tz}
for a detailed discussion): starting as a hot and dense intermediate
state eventually with ``free'' ({\it i.e.}, only partly or weakly bound)
quarks and gluons the system cools down until the exchange of
resonance species stops (chemical freeze-out) and finally the
particles do not scatter any longer elastically (thermal
freeze-out). Our goal is, however, to gather information of the early
phase, {\it i.e.\ }before the chemical freeze-out occurs.

One of the main probes (if not the only one) which carry undistorted
information over the entire history are di-leptons ($e^+e^-$ or $\mu^+\mu^-$)
as they are not hampered by final-state interactions. This included
the first-chance radiation (bremsstrahlung), radiation from
short-lived resonances (the real ``messengers'' which have a life time
shorter then those of the fireball) and post-freeze-out sources, such as
the late $\pi^0$ and $\eta$ Dalitz decays, which can be subtracted if their
yields and distributions at the freeze-out point are precisely known.
Now, after a successful decade of measurements of electromagnetic
probes (summarized in~\cite{Tserruya:2009zt,Rapp:2009yu}) there is
common agreement that only by systematic studies in various systems
over a  wide energy range, conclusions of the following questions can
be drawn: when does the onset of deconfinement appear, and how it is
related to chiral symmetry restoration?

Recently, HADES~\cite{nim} has added valuable measurements of
di-electron mass spectra for light $A+A$~\cite{hades_papers}
systems as well as for (quasi-free) $p+N$
collisions~\cite{hades_nn_papers} at kinetic beam energies
of 1-2~AGeV and 3.5~GeV, which has triggered a lot of theoretical
activities~\cite{hades_theo}.  One of the main features of HADES is that at the
same time it measures also hadrons: pions for absolute
normalization~\cite{hades_pions} and hadrons containing
strangeness~\cite{hades_strange}.  

To summarize the setup at this
point, HADES is a magnetic spectrometer, consisting of up to 4 planes
of Mini Drift Chambers (MDC) with a toroidal field
created by a superconducting magnet.  Particle
identification is based on momentum and time-of-flight measurements.
In addition, a Ring Imaging Cherenkov detector (RICH) and an
electromagnetic Pre-Shower detector provide electron identification
capabilities.

HADES will continue its program at its current place at SIS-18, and then move to
the upcoming FAIR
%\footnote{FAIR: ``International Facility for Antiproton and
%  Ion Research''} 
accelerator complex. Here, HADES will continue its
experimental program up to kinetic beam energies per nucleon of 8~GeV at SIS-100.
This is one of the main reason for upgrading the HADES
detector and its trigger and readout system, which will be outlined 
in the following.

\section{The upgrade program for HADES-at-FAIR}

\subsection{Motivation}

\begin{figure}[t]
  \begin{center}
    \includegraphics[width=0.7\columnwidth]{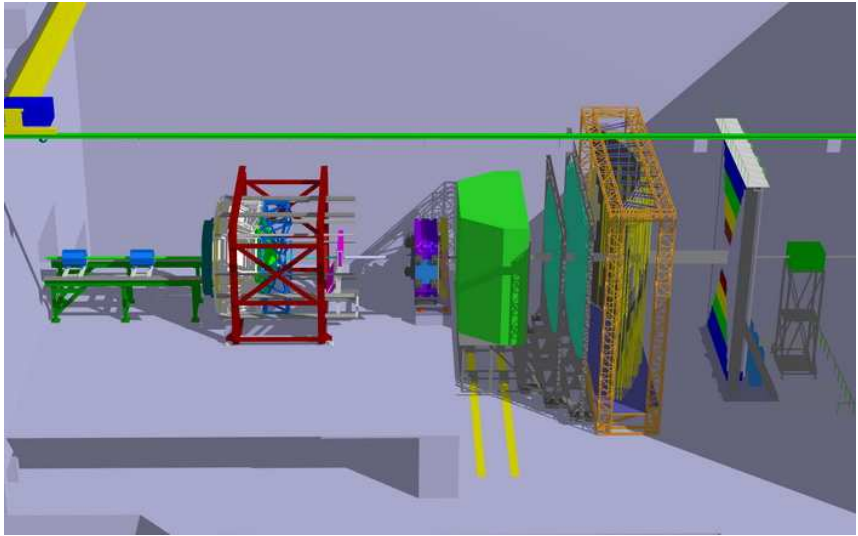}
    \caption{\label{fig:hades_cbm} Artistic view of the HADES
      spectrometer (left) among with CBM (right) in the new cave (courtesy of W. Niebur, GSI).  }
  \end{center}
\end{figure}

In the kinetic beam energy range of 2-40~AGeV, the di-lepton landscape
is completely ``terra incognita''.  At energies starting from 8 GeV,
the CBM experiment~\cite{cbm} will measure di-electrons and
di-muons. HADES, on the other hand, will provide the link between 2-8
GeV. Therefore the setup will be moved into the CBM cave and share the
same beam line (see fig.~\ref{fig:hades_cbm}).

The advantage of re-using the HADES setup is that is has all required
features for di-electron measurements (as well as strangeness), and a
higher acceptance as CBM for energies up to 8~AGeV. However, it needs a
larger granularity (realized by the addition of the new RPC detector
(``Resistive Plate Chamber''), discussed elsewhere~\cite{rpc}), better
tracking and a slightly modified RICH detector.  Furthermore, the
large data volumes, expected in experiments with heavy collision
systems (Au+Au) already at SIS18 and with higher energies at the new
FAIR facility, require bandwidths which cannot be achieved by the
current system.  Since the currently used data acquisition system was
designed ten years ago, it was reasonable to reconsider the whole
concept and make use of new technologies.

In addition, higher statistics would allow for the multi-differential
measurement of additional degrees of freedom of the virtual photon
($\gamma^* \to e^+e^-$): As the virtual photon is not mass-less it
has two polarization quantities, which are sensitive to the hadronic
source and its possible polarization.  One of them (the longitudinal
polarization), which becomes manifest in the helicity angle
$\theta^{ee}_e$, can be used for nailing down the composition of the cocktail~\cite{brat}
and the nature of any additional excess~\cite{na60_ani}.

Two other parameters are the emission angles of the virtual photon.
Here, the discussion of the azimuthal angle makes only sense in the
context of a measurement of the reaction plane.

\subsection{Forward wall}

Future HADES experiments at FAIR will therefore profit from a Forward
Wall (FW) consisting of 288 plastic scintillator cells, which has
been added to the HADES setup in 2007. The FW was already used in 2007
to detect spectator protons in the break-up of projectile deuterons
with an incident energy of 1.25 AGeV~\cite{hades_nn_papers}.
The FW, consisting of 140 small cells (4x4~cm$^2$) close to the
symmetry axis, 64 middle cells (8x8~cm$^2$) and 84 big cells (16x16~cm$^2$), 
is positioned seven meters
downstream from the target.

The occupancy of single cells has been studied using UrQMD simulations~\cite{urqmd}
for semi-central Au + Au collisions at 1.0 AGeV compared to those at 8.0
AGeV; it stays below 0.5 in all cases. It should also be noted that UrQMD
provides only spectator nucleons and no fragments, {\it i.e.\ }it overestimates the
nucleon multiplicity, affecting mostly the inner part with 20~cm radius.
The occupancy increases only about a factor two going from SIS18 to
low FAIR energies. The reaction-plane angular resolution
$\sigma(\phi)$ deduced from proton hits in the FW is about 48 degrees.

This new, already working detector allows 
to measure the reaction plane, study elliptic flow and
to define the relative azimuthal emission angle $\phi$ of the virtual photon.

\subsection{Calorimeter for HADES}

Di-electron results obtained at SPS and RHIC demonstrate a large pair
excess in the intermediate ($0.14 < M < 0.6$~GeV/c$^2$) mass region. An accurate
determination of this excess depends on precise knowledge of
the hadronic cocktail, which for 2-40~AGeV is dominated in this mass
region by the $\eta$ Dalitz decay.  Furthermore, a convenient normalization
of the di-electron spectra is naturally given by the $\pi^0$ yield. 

For the SIS-18 energy range, the production of neutral mesons has been
studied extensively by the TAPS collaboration via photon
calorimetry. However, for the 2-40 AGeV range no data at all do
presently exist, with the consequence that any interpretation of
future di-electron data ({\it i.e.\ } the subtraction of 
the long-lived components) would have to depend solely on theoretical
models, {\it e.g. }transport calculations or appropriate hydrodynamical
models. In order to remedy this situation we propose to measure the
respective $\pi^0$ and $\eta$   meson yields together with the di-electron
data. This can be achieved by replacing the HADES Pre-Shower detector,
located at forward angles ($18^\circ < \theta < 45^\circ$ ), 
with an electromagnetic
calorimeter (ECAL).

The option, which is currently under investigation, is to recuperate a
lead-glass calorimeter from the former OPAL experiment at LEP and
adapt it to HADES. The existing OPAL modules have a geometry of
9.4x9.4 cm$^2$, 840 modules in total. 

\subsection{The new trigger and readout system}

For the data rate of the new DAQ system, 100~kHz are aimed for $N+N$
collisions, and 20~kHz for heavy systems (Au+Au at 2~AGeV, Ni+Ni at
8~AGeV).  To fulfill this goal, a new trigger and readout board
(TRB), a multi-purpose electronic device with on-board DAQ
functionality, has been developed~\cite{trb}.  A first version of this hardware,
has already been used to read out the HADES
forward wall mentioned above.

\begin{figure}[t]
  \begin{center}
    \includegraphics[width=0.7\columnwidth]{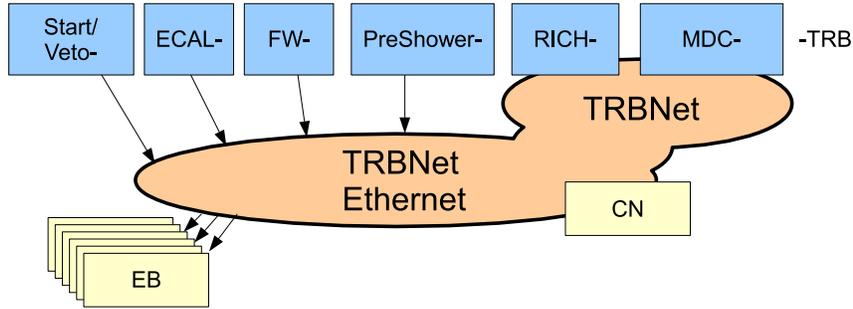}
    \caption{\label{fig:over} Sketch of the new trigger and readout system.  }
  \end{center}
\end{figure}

The readout electronics is on board and the data are transported to
mass storage via 100~Mbit/s Ethernet and the UDP internet
protocol. Optional time measurements are done using a 124-channel Time
to Digital Converter (TDC) device based on the HPTDC~\cite{hptdc-ref}. As
the concept has been tested successfully during HADES data taking
periods, its design has been extended to serve as the new standard
readout module for all HADES sub-detectors by adding a high-speed
connector and pluggable dedicated detector-specific add-on
modules~\cite{palka}.  But also the trigger distribution will be
replaced by a tree-like trigger distribution via optical TRB-Hubs with
only point-to-point links, realized with 2 Gbit/s optical links, which can
also be used for the transport of the data. The planned
computing nodes~\cite{cn} (CN), the same type which is
going to be used for PANDA~\cite{panda}, providing high-speed pattern
recognition, can easily be integrated. 
For the readout of the drift chambers 
a TRB-add-on has been made which contains 32 optical transceivers thus replacing
the electrical connection which turned out to be a source of noise.
On the front-end side, the counterpart, we have developed 
a very small, highly integrated optical end point driver card.

For easy monitoring and operation of the large number of modules needed,
the TRBNet, a new hardware-independent network protocol has been developed.
It allows the simultaneous (interleaved) distribution of trigger information,
which has to be done with a low latency, and the transportation
of large amounts of data using only one data link. Different logical channels
provide priority levels, while buffers on each connection guarantee protection
from data loss. Finally, multiple event builders (EB) are connected via
Ethernet to the TRBNet.  This concept is visualized in
fig.~\ref{fig:over}.

\section{Summary and outlook}

In summary, we presented the concept of the ``HADES-at-FAIR'' project.
The HADES spectrometer, after finishing the currently ongoing upgrade, is ready
for experiments at the future SIS-100 synchrotron of the FAIR facility.
We expect to fill the knowledge gap between 2-8~AGeV and provide first data at the very first day
once the beam is injected into the CBM cave.

\section*{Acknowledgments}

\renewcommand{\baselinestretch}{0.9}
\footnotesize
The collaboration gratefully acknowledges the support by BMBF grants
06MT238, 06GI146I, 06FY171, and 06DR135, by the VI-146 
and the Helmholz Alliance Programm HA216/EMMI ``Extremes
of Density and Temperature: Cosmic Matter in the Laboratory''
of the Helmholtz
Association, by the DFG cluster of excellence EClust 153 and by
Maier-Leibniz-Laboratorium Munich (Germany),
by GSI (TM KRUE,GI/ME3,OF/STR),
by grants 
MSMT LC07050, LA316 and
GAASCR IAA100480803 (Czech Republic),
by grant KBN 1P03B 056 29 (Poland),
by INFN (Italy), 
by CNRS/IN2P3 (France), 
by grants FPA2006-09154, PGIDIT06PXIC296091PM and CPAN:CSD2007-00042
(Spain),
by grant POCI/FP/81982/2007 (Portugal),
by grant UCY-10.3.11.12 (Cyprus),
by INTAS grant 06-1000012-8861 and by EU contract RII3-CT-2004-506078.
\\
This work was supported by the Hessian LOEWE initiative through the
Helmholtz International Center for FAIR (HIC for FAIR).
\renewcommand{\baselinestretch}{1}
\normalsize

% For Figures insertion uncomment the next section

%\begin{figure}
%\includegraphics{figurename}
%\caption{Your caption here}
%\label{fig01} % optional figure label, must be unique
%\end{figure}

%%%%%%%%%%%%%%%%%%%%%%%%%%%%%%%%%%%%%%%%%%%%%%%%%%%%%%%%
% End of the paper
%


\begin{thebibliography}{9}

\bibitem{Hayano:2008vn}
  R.~S.~Hayano and T.~Hatsuda,
  %``Hadron properties in the nuclear medium,''
  arXiv:0812.1702 [nucl-ex].
  %%CITATION = ARXIV:0812.1702;%%

\bibitem{BraunMunzinger:2008tz}
  P.~Braun-Munzinger and J.~Wambach,
  %``The Phase Diagram of Strongly-Interacting Matter,''
  arXiv:0801.4256 [hep-ph].
  %%CITATION = ARXIV:0801.4256;%%

\bibitem{Tserruya:2009zt}
  I.~Tserruya,
  %``Electromagnetic Probes,''
  arXiv:0903.0415 [nucl-ex].
  %%CITATION = ARXIV:0903.0415;%%


\bibitem{Rapp:2009yu}
  R.~Rapp, J.~Wambach and H.~van Hees,
  %``The Chiral Restoration Transition of QCD and Low Mass Dileptons,''
  arXiv:0901.3289 [hep-ph].
  %%CITATION = ARXIV:0901.3289;%%

\bibitem{nim}
  G.~Agakishiev {\it et al.}  (HADES),
  %``The High-Acceptance Dielectron Spectrometer HADES,''
  arXiv:0902.3478 [nucl-ex].
  %%CITATION = ARXIV:0902.3478;%%

\bibitem{hades_papers}
G.~Agakishiev {\it et al.}  (HADES),
  %``Study of dielectron production in C+C collisions at 1 AGeV,''
  Phys.\ Lett.\  B {\bf 663} (2008) 43
  [arXiv:0711.4281];
  %%CITATION = PHLTA,B663,43;%%
G.~Agakichiev {\it et al.}  (HADES),
  %``Dielectron production in C-12 + C-12 collisions at 2-AGeV with HADES,''
  Phys.\ Rev.\ Lett.\  {\bf 98} (2007) 052302
  [arXiv:nucl-ex/0608031].
  %%CITATION = PRLTA,98,052302;%%

\bibitem{hades_nn_papers}
K.~Lapidus {\it et al.}  (HADES),
  %``Dielectron production in pp and dp collisions at 1.25 GeV/u with HADES,''
  arXiv:0904.1128 [nucl-ex].
  %%CITATION = ARXIV:0904.1128;%%

\bibitem{hades_theo}
K.~Schmidt, E.~Santini, S.~Vogel, C.~Sturm, M.~Bleicher and H.~St\"ocker,
  %``Production and evolution path of dileptons at HADES energies,''
  arXiv:0811.4073 [nucl-th];
  %%CITATION = ARXIV:0811.4073;%%
E.~Santini, M.~D.~Cozma, A.~Faessler, C.~Fuchs, M.~I.~Krivoruchenko and B.~Martemyanov,
  %``Dilepton production in heavy-ion collisions with in-medium spectral
  %functions of vector mesons,''
  Phys.\ Rev.\  C {\bf 78} (2008) 034910
  [arXiv:0804.3702];
  %%CITATION = PHRVA,C78,034910;%%
E.~L.~Bratkovskaya and W.~Cassing,
  %``Dilepton production and off-shell transport dynamics at SIS energies,''
  Nucl.\ Phys.\  A {\bf 807} (2008) 214
  [arXiv:0712.0635].
  %%CITATION = NUPHA,A807,214;%%

\bibitem{hades_pions}
G.~Agakishiev {\it et al.}  (HADES),
Eur. Phys. J {\bf  A 40} (2009) 45
  %``Measurement of charged pions in 12C + 12C collisions at 1A GeV and 2A GeV
  %with HADES,''
  [arXiv:0902.4377];
  %%CITATION = ARXIV:0902.4377;%%
P.~Tlusty {\it et al.}  [The HADES Collaboration],
  %``Charged pion production in C+C and Ar+KCl collisions measured with HADES,''
  arXiv:0906.2309 [nucl-ex].
  %%CITATION = ARXIV:0906.2309;%%


\bibitem{hades_strange}
G.~Agakishiev {\it et al.}  (HADES),
  %``Phi decay: a relevant source for K- production at SIS energies?,''
  arXiv:0902.3487 [nucl-ex].
  %%CITATION = ARXIV:0902.3487;%%


\bibitem{cbm}
  V. Friese,
  %``The CBM experiment at GSI/FAIR,''
  Nucl.\ Phys.\  A {\bf 774} (2006) 377;
  %%CITATION = NUPHA,A774,377;%%
  P. Senger {\it et al.},
  %``Cbm At Fair,''
  PoS  {\bf CPOD2006} (2006) 018;
  %%CITATION = POSCI,CPOD2006,018;%%
  V. Friese,
  %``The CBM experiment at FAIR,''
  PoS  {\bf CPOD07} (2007) 056 (CBM).
  %%CITATION = POSCI,CPOD07,056;%%

\bibitem{rpc}
A. Blanco {\it et al.} (HADES), Nucl. Instr. and Meth. {\bf A 602} (2009) 691.

\bibitem{brat}
E.~L.~Bratkovskaya, W.~Cassing and U.~Mosel,
  %``Probing hadronic polarizations with dilepton anisotropies,''
  Z.\ Phys.\  C {\bf 75} (1997) 119
  [arXiv:nucl-th/9605025];
  %%CITATION = ZEPYA,C75,119;%%
E.~L.~Bratkovskaya, W.~Cassing and U.~Mosel,
  %``Dilepton anisotropy from $p + Be$ and $Ca + Ca$ collisions at BEVALAC
  %energies,''
  Phys.\ Lett.\  B {\bf 376} (1996) 12
  [arXiv:nucl-th/9601018];
  %%CITATION = PHLTA,B376,12;%%
T.~I.~Gulamov, A.~I.~Titov and B.~K\"ampfer,
  %``Asymmetry of the dielectron emission rate in an isospin - asymmetric pion
  %medium,''
  Phys.\ Lett.\  B {\bf 372} (1996) 187;
  %%CITATION = PHLTA,B372,187;%%
A.~I.~Titov and B.~K\"ampfer,
  %``In-medium modification and decay asymmetry of omega mesons in cold nuclear
  %matter,''
  Phys.\ Rev.\  C {\bf 76} (2007) 065211
  [arXiv:0709.1393].
  %%CITATION = PHRVA,C76,065211;%%


\bibitem{na60_ani}
R.~Arnaldi {\it et al.}  (NA60),
  %``First results on angular distributions of thermal dileptons in nuclear
  %collisions,''
  arXiv:0812.3100 [nucl-ex].
  %%CITATION = ARXIV:0812.3100;%%


\bibitem{urqmd}
M.~Bleicher {\it et al.},
  %``Relativistic hadron hadron collisions in the ultra-relativistic quantum
  %molecular dynamics model,''
  J.\ Phys.\ G {\bf 25} (1999) 1859
  [arXiv:hep-ph/9909407].
  %%CITATION = JPHGB,G25,1859;%%

\bibitem{trb}
I.~Fr\"ohlich {\it et al.} (HADES),
  %``A General Purpose Trigger and Readout Board for HADES and
  %FAIR-Experiments,''
  IEEE Trans.\ Nucl.\ Sci.\  {\bf 55} (2008) 59.
  %%CITATION = IETNA,55,59;%%

\bibitem{hptdc-ref} M. Mota {\it et al.}
%  ``A flexible multi-channel high-resolution time-to-digital converter ASIC'',
  2000 IEEE Nuclear Science Symposium
  Conference Record, Volume 2, 9/155~-~9/159
  (2000). 

\bibitem{palka}
 M. Palka  {\it et al} (HADES),
NSS 2008 conference record, IEEE (2008) 1398. 

\bibitem{cn} 
W.~K\"uhn {\it et al.}  (PANDA),
  %``Fpga Based Compute Nodes For High Level Triggering In Panda,''
  J.\ Phys.\ Conf.\ Ser.\  {\bf 119} (2008) 022027.
  %%CITATION = 00462,119,022027;%%

\bibitem{panda} 
J. G. Messchendorp (PANDA),
  %``Hadron Physics with Anti-Protons: The PANDA Experiment at FAIR,''
  arXiv:0711.1598 [nucl-ex];
  %%CITATION = ECONF,C070910,123;%%

\end{thebibliography}
\end{document}